# The Diversity Paradox: How Nature Resolves an Evolutionary Dilemma

*James M. Whitacre*[*,‡] *and Sergei P. Atamas*[†]

From the [*]CERCIA Computational Intelligence Lab, University of Birmingham, U.K., and [†]Departments of Medicine and Microbiology & Immunology, University of Maryland School of Medicine, and Baltimore VA Medical Center, Baltimore, MD, U.S.A.

[‡]Corresponding author

James M. Whitacre, PhD, School of Computer Science, University of Birmingham, Edgbaston, B15 2TT, United Kingdom; E-mail: jwhitacre79@gmail.com

Sergei P. Atamas, MD, PhD, University of Maryland School of Medicine, 10 South Pine St., MSTF 8-34, Baltimore, MD 21201, U.S.A.; E-mail: satamas@umaryland.edu

**Insight Statement:** Cryptic Genetic Variation (CGV) and degeneracy paradigms for adaptation are integrated in this review, revealing a common set of principles that support adaptation at multiple levels of biological organization. Using simulation studies, molecular-based experimental systems, and principles from population genetics, we demonstrate that CGV and degeneracy reflect complementary top-down and bottom-up, respectively, conceptualizations of the same phenomenon and capture a universal feature of biological adaptive processes. Degeneracy's role in resolving conflicts between stabilizing selection and directional selection has numerous applications. For instance, we consider improved measures of intraspecific biodiversity for conservation planning, adaptive therapeutic strategies for defeating evolvable pathogens and aggressive cancers, and the translation of degeneracy as a design principle for improving flexibility in a number of abiotic technologies.

## Abstract

Adaptation, through selection, to changing environments is a hallmark of biological systems. Diversity in traits is necessary for adaptation and can influence the survival of a population faced with environmental novelty. In habitats that remain stable over many generations, effective stabilizing selection reduces quantitative trait differences within populations, thereby appearing to remove the diversity needed for heritable adaptive responses in new environments. Paradoxically, field studies have documented numerous populations under long periods of stabilizing selection and evolutionary stasis that have rapidly evolved under changed environmental conditions. In this article, we review how cryptic genetic variation (CGV) resolves this diversity paradox by allowing populations in a stable environment to gradually accumulate hidden genetic diversity that is revealed as trait differences when environments change. Instead of being in conflict, environmental stasis supports CGV accumulation and thus appears to facilitate rapid adaptation in new environments as suggested by recent CGV studies. A phenomenon similar to CGV, known as degeneracy, has been found to similarly support both genetic and non-genetic adaptation at many levels of biological organization. Degenerate, as opposed to diverse or redundant, ensembles appear functionally redundant/interchangeable in certain environmental contexts but functionally diverse in others. CGV and degeneracy paradigms for adaptation are integrated in this review, revealing a common set of principles that support adaptation at multiple levels of biological organization. Though a discussion of simulation studies, molecular-based experimental systems, principles from population genetics, and field experiments, we demonstrate that CGV and degeneracy reflect complementary top-down and bottom-up, respectively, conceptualizations of the same basic phenomenon and arguably capture a universal feature of biological adaptive processes.

A greater appreciation of degeneracy's role in resolving conflicts between current (stabilizing) selection and future changes in (directional) selection should have important practical implications. In conservation biology for instance, CGV revealed under predicted stress conditions should provide a more adaptively significant measure of intraspecific biodiversity for conservation planning. Similar principles could guide the development of new therapeutic strategies for defeating evolvable pathogens and preventing the evolution of therapy resistance in cancer. In aggressive cancers for instance, a stable tumour environment enables CGV to gradually accumulate and may enable the rapid evolution of therapy resistance towards targeted drugs. On the other hand, the introduction of a dynamic therapeutic strategy should eliminate CGV over time and create population bottlenecks that reduce the drug-resistance adaptation potential of a cancer cell population. While CGV is a purely biological property, degeneracy can be defined and measured within any system – biological or non-biological –comprised of functionally versatile elements. We also review recent efforts to apply the degeneracy concept in addressing exploitation-exploration conflicts that arise in various social and technological systems. Results are thus far promising and suggest that key determinants of flexibility in biological systems can be translated into design principles for improving adaptive capabilities within non-biological contexts.

Draft: Submitted to Integrative Biology (2011)

## Background

Various factors can contribute to the maintenance of heritable trait differences in a population, including balancing selection (e.g., frequency dependent selection), gene flow between populations, assortative mating, and mutation-selection balance. Despite the contributions from these factors, populations often appear relatively homogeneous in most traits, i.e. the variance in quantitative traits is small compared to the change in mean trait values that can occur following periods of rapid adaptation [1]. In habitats that remain stable over many generations, stabilizing selection drives many populations to converge towards the most adaptive trait values. In principle, such phenotypic homogeneity should limit a population's ability to successfully adapt to novel environmental stresses.

While many habitats appear stable, all habitats eventually change.[1] There is an inherent conflict between the stable habitat conditions that drive a population to converge to the most adaptive traits on the one hand, and the need to maintain diversity and display heritable trait differences when faced with environmental novelty on the other hand. Stated more abstractly, there is an evolutionary conflict between the immediate need to exploit current conditions and the longer-term need to diversify (bet-hedging) in order to adapt to future unexpected events. Similar exploration-exploitation conflicts arise within a number of socially and economically relevant non-biological systems that are subjected to variation and selection in a dynamic environment.

Exploration-exploitation conflicts are possibly best illustrated within population-based dynamic optimization research [2]. Within this sub-field of operations research, evolutionary algorithms are employed where solution parameters to an optimization problem represent a genotype and the corresponding solution performance on an optimization problem's objective function represents fitness. Using a population of these solutions, the more fit solutions are preferentially mated (recombination of solution vectors) and mutated (perturbation of solution vector within solution space) to generate new offspring solutions that are selectively bred in the next generation [3]. With a static objective function, populations consistently converge over many generations to eventually display low variance in population fitness and low genetic diversity [4]. However the more that the population converges, the slower it adapts to changes in the objective function, thereby limiting algorithm performance on dynamic optimization problems. This conflict between exploiting current problem conditions and maintaining diversity to prepare for future problem changes has been addressed in this field using numerous tools that subvert fitness-biased selection or otherwise encourage trait diversity to be maintained. However, by enforcing diversity, this limits the speed and extent that the simulated population can adapt to the current problem definition, thus revealing a fundamental tradeoff between short-term and long-term algorithm performance.

Understanding the mechanisms by which natural populations are able to resolve this "diversity paradox" could provide insights for reconciling similar conflicts in operations research [4] as well as exploration-exploitation conflicts in strategic planning [5], systems engineering [5], and peer review [6]. Moreover, and as we elaborate on later in the article, these issues are widely relevant to ecological conservation efforts and the eradication of evolvable pathogens.

---

[1] Factors such as migration and habitat tracking reduce the frequency, but not the inevitability, of experiencing this change.

## How Nature Resolves the Diversity Paradox

**Cryptic Genetic Variation**: Natural populations appear to resolve the diversity paradox through a phenomenon known as cryptic genetic variation (CGV). In a stable environment, CGV is hidden, or cryptic, with organism phenotypes remaining unchanged despite genetic mutation [16]. Such mutational robustness –the stability of the phenotype to genetic mutations– was originally predicted to impede evolution [7, 8] because it lowers the number of distinct heritable phenotypes that are mutationally accessible from a single genotype and it reduces selective differences within a genetically diverse population [8]. There is still confusion about the role of CGV in evolution, due to long-standing difficulties in understanding relationships between the robustness and the adaptive modification of traits. Only in the last decade have arguments been put forth to explain how evolution can be supported by mutational [8-10] and environmental [11] forms of robustness. Mutational robustness is believed to support evolution in two ways: (i) in a stable environment it establishes fitness-neutral regions in fitness landscapes from which large numbers of distinct heritable phenotypes can be sampled (via genetic mutations that lead to genotypes that are not members of the neutral set) [8, 10] and (ii) it allows for cryptic genetic differences to accumulate in a population with subsequent trait differences revealed in an environment-dependent manner [11].

Under stabilizing selection, individuals become more phenotypically similar, yet can accumulate (through selectively neutral mutations) genetic differences that have the potential to be revealed as trait diversity should the environment change and directional selection emerge. CGV thus resolves the diversity paradox by providing the heritable trait diversity that is necessary for adaptation under new stressful environments, while bypassing the negative selection that limits phenotypic variability under stable conditions. Although CGV revealed by the environmental change has long been implicated as a pathway for adaptation [12-15], its role in resolving tensions between stabilizing and directional selection is not widely recognized. Importantly, by enabling the accumulation of CGV, stable environments may actually support a population's ability to rapidly adapt in new environments. Given the punctuated dynamics of ecosystem regime shifts [16] and the corresponding changes that often occur to habitat range and ecological opportunities, adaptation capabilities afforded by CGV may provide an essential ingredient for the persistence of life in a changing yet meta-stable environment.

Recent studies support this role for CGV in adaptation with some indications that environment-exposed CGV is fixed in populations at least as often as adaptations initiated by novel alleles [17]. In a recent study by Wagner and colleagues using a ribozyme RNA enzyme as their experimental model, they confirmed that ribozyme populations containing cryptic variation adapt more rapidly to new substrates compared to ribozyme populations that did not contain cryptic mutations [14]. Another recent study of thre espine stickleback by McGuigan *et al.* indicates that CGV enables populations experiencing evolutionary stasis to be poised for rapid adaptation in new environments [15]. In that study, they selected a species that previously underwent rapid speciation when local populations were introduced to a new habitat. These adaptations were associated with readily observable trait changes that could be directly linked to survival and reproductive success within the new habitat. In their experiments, offspring from the original species were bred under original and new habitat conditions. When raised in the original habitat, offspring developed similar traits, while in the new habitat trait variations were observed that corresponded with those that are beneficial to the new species. Because both populations were bred in artificially controlled homogeneous environments, the observed trait variations could readily be attributed to an environment-exposed release of CGV. In other words, a species in evolutionary stasis with few observable trait differences was shown to undergo rapid evolution using cryptic genetic variation.

**Degeneracy**: CGV is similar to a phenomena observed in molecular and systems biology that are generalized under the term of degeneracy. Degenerate, as opposed to diverse or redundant, ensembles appear functionally redundant in certain environmental contexts but functionally diverse in others [18, 19]. Such context-dependent similarity in functions/traits among diverse units of an ensemble, and, reciprocally, context-dependent dissimilarity of redundant units, is common in biology. It can be observed at the molecular and cellular levels of gene regulation, in proteins of every functional class (e.g. enzymatic, structural, or regulatory) [20], in protein complex assemblies [21]; also in ontogenesis (see page 14 in [22]), the nervous system [23], metabolic pathways [24], and in cell signaling [25].

Degeneracy is defined relationally to diversity and redundancy, by comparing the behavior of units under various environmental contexts. For degeneracy to exist, the units being studied must exhibit functional versatility which, depending on the context, may manifest as enzymatic substrate ambiguity, multiple ligand-receptor cross-reactivity, protein moonlighting, or multiple use of organs, e.g., use of fins for swimming and crawling. Functional versatility implies that the units can change their behavior in a manner that is revealed by changes in the local environment. Then, if degeneracy exists amongst two functionally versatile units, this means that there will be contexts in which the two units will display behaviors that appear to be functionally the same and other contexts where the two will appear to be functionally distinct [26]. Degenerate units will manifest diverse, yet functionally overlapping, behavior that is versatile and can be co-opted within novel environments to occasionally display new functional relevance. As we highlight through selected examples, this functional versatility fundamentally underpins degeneracy and exaptation in complex biological systems.

**CGV is a case of adaptive degeneracy:** Degeneracy is also observed in natural populations and demes. Physiological, immunological, cognitive, behavioral, and even morphological traits demonstrate performance versatility over numerous environmental backgrounds. Furthermore, these traits can appear very similar across a population that is placed in its native environment, and yet selectively relevant differences in these traits can be revealed when the population is presented with novel stresses. The heritable component of such cryptic trait differences is aptly referred to as cryptic genetic variation. Gibson defines CGV as "standing genetic variation that does not contribute to the normal range of phenotypes observed in a population, but that is available to modify a phenotype that arises after environmental change…" [27]. Thus, by definitions of CGV and degeneracy, CGV represents a heritable form of phenotypic degeneracy in populations, and as proposed here, exemplifies a more general phenomena for resolving exploitation-exploration conflicts in natural, social and technological systems.

It is important to note that CGV describes a relationship between genotypic and phenotypic variation, and is not solely a genetic property such as genetic diversity. In populations with CGV, the interactions between individual genotypes and the environment (GxE interactions) are such that organisms can appear phenotypically similar in some environments but phenotypically distinct in others. Importantly, CGV, GxE, and epistatic interactions provide statistical descriptions of the relationship between genotype, phenotype and environment. In contrast, degeneracy describes an innate versatility in the characters that make up a phenotype and whose selective value can be revealed in the right environment. In this way, degeneracy and CGV provide bottom-up and top-down, respectively, descriptions of the same biological phenomena.

**Degeneracy and evolution**: Degeneracy facilitates adaptation at many levels of biological organization [19, 20, 23, 28]. In some cases, degeneracy facilitates adaptive robustness through the provision of functional redundancy [25, 29-31]. In other cases, degeneracy provides functional diversity and enables adaptive phenotypic change as seen with CGV in natural populations. The adaptive immune response of T cells provides another illuminating example of degeneracy-driven adaptation that is useful for integrating, yet also distinguishing between, degeneracy and CGV paradigms [32, 33]. Naïve T cells typically display similar (inactive) phenotypes under "normal" conditions (devoid of antigens) but can also reveal large phenotypic differences when presented with novel (antigen) environments that in turn drive rapid clonal expansion. Because these environment-revealed trait differences are heritable through genetic differences in the alpha/beta chain segments of the T cell receptor (TCR), this cell population is thereby poised to rapidly evolve using cryptic genetic variation.

It would seem then that CGV adequately explains this adaptive response and there is no need to confuse the discussion by mentioning degeneracy. However, proper functioning of the adaptive immune system is vitally dependent on binding ambiguity between TCR and MHC-antigen complexes found on the surfaces of professional Antigen Presenting Cells (APCs). This promiscuity provides each TCR with sufficient affinity to drive T cell activation in response to numerous distinct antigens. Without this promiscuity, impossibly large TCR repertoires (and T cell populations) would be needed in order to effectively cover the antigenic space [34]. In other words, TCR ambiguity allows the population to be capable of adapting to a much wider range of antigenic contexts. On the other hand, partial overlap in TCR affinity also allows the T cell population to invoke similar responses under conditions devoid of antigens. In particular, similarities facilitate non-trivial cell-cell adhesion events between T cells and APCs that enable scanning for cell surface antigens with inactivation under normal conditions. In short, it is the conditional similarities and differences (degeneracy) in the versatile TCR repertoire that enable rapid adaptation within a changing antigenic environment. More generally, degeneracy can facilitate rapid adaptation because some of the elements in the repertoire are already pre-adapted to some degree and can be elaborated upon over several generations to enhance functionality and fitness. If elements in the repertoire exhibited extreme functional specificity that relied on precise environmental conditions then adaptation would be highly unlikely and could only occur by chance. These basic relationships between versatility and exaptation (cf [35]) are not limited to the adaptive immune response but are fundamental attributes of adaptive responses throughout many biological systems [23].

As a paradigm for adaptation, degeneracy principles may also be relevant to non-biological systems such as social [6] and technological systems [36] because degeneracy will arise amongst any set of versatile components that modify their function in response to a variety of contexts. Unlike high specificity components whose functionality is limited to a very specific context, degenerate components are likely to exhibit a change in function that is differentially expressed across a degenerate repertoire. Many of these changes in function are likely to be maladaptive; however, those that are not provide useful information about where additional beneficial adaptations are likely to be found. This has been best described at the level of protein evolution where proteins can maintain one functionally relevant conformation while occasionally sampling other conformational structures that are sometimes co-opted for new useful functions in new environments [37, 38]. Functional versatility and degeneracy principles have been well characterized at a molecular level, however these same principles equally describe environment-revealed differences in high level versatile traits involving adaptive foraging, nest-building, and predator avoidance.

Hypothetical populations with high trait diversity but no degeneracy will not harbor characters that flexibly respond to environmental change, and thus any beneficial trait discovered must constitute a chance encounter between a highly specific genotype-environment pairing. In theoretical biology, these points of difference between degenerate and non-degenerate systems can be related to rugged and smooth fitness landscapes. In rugged landscapes, adaptations occur by chance alone while in the latter, genotypic and fitness changes are correlated such that the benefit of a genotype suggests a non-negligible likelihood of additional beneficial mutations nearby. In theoretical simulations, we have shown that only fitness landscapes with neutrality created by degeneracy will provide access to the phenotypic diversity needed to enhance a population's evolutionary potential [4, 10, 39]. From these theoretical perspectives, degeneracy thus appears to contribute to a mechanistic foundation for realizing the Darwinian principles of heritable variation and selection [19, 23, 40].

## Experimental Principles for Evaluating the role of CGV and Degeneracy in Evolution

There are a number of existing experimental and analytical tools that are relevant for exploring the role of CGV in evolution and that involve the analysis of standing genetic variation (reviewed in [41]). For instance, alleles fixed from standing genetic variation should have a different signature of selection compared to *de novo* mutations. In particular, since multiple copies of a selectively neutral allele can grow in a population through genetic drift, recombination will reduce the strength of genetic hitchhiking (lower linkage disequilibrium) at all but the nearest loci creating a narrower valley of low polymorphism that surrounds a CGV locus in comparison to selective sweeps on *de novo* mutations, see [41]. Confounding factors, such as population growth and non-random mating, can, however, provide alternative explanations for these selection signatures [41].

An alternative approach is to infer CGV-driven adaptation by looking for beneficial alleles in a new environment that are present as standing variation in the ancestral population. The phylogenetic history of alleles provides similar evidence for CGV-driven adaptation events [41] and other approaches may also be possible [42]. Although these approaches may determine whether standing genetic variation was a primary contributor to adaptation, they do not address the hide and release of the corresponding traits. To demonstrate a specific role for CGV, time intensive experiments where environmental manipulations reveal previously hidden adaptive phenotypes appear to remain an important experimental tool, e.g. see [15].

## The Implications of Degeneracy in Evolution

The diversity paradox exists because natural selection in stable environments appears to continually remove the very heritable diversity in traits that is later needed when changes in selection take place. Phenotypic degeneracy in general, and CGV in particular, might resolve this paradox because diverse cryptic changes can accumulate in a population without noticeable phenotypic effects yet can ultimately provide the trait differences necessary for adapting to future environmental novelty. With degeneracy appearing to be ubiquitous at all levels of biological organization [23], and heritable through CGV, a growing number of researchers are viewing degeneracy as a major contributor and universal facilitator of evolution [23, 40]. Here we briefly explore how the degeneracy concept could provide new insights into evolutionary processes that arise in a variety of biological and non-biological domains.

**Conservation Biology:** One important example is seen in evolutionary ecology where the relevance of CGV is not widely appreciated. In conservation biology, intraspecific genetic diversity is a well known factor in extinction risk. However, only genetic variants that reveal trait differences under the environmental stresses encountered can mitigate extinction risks. It should be possible to quantify GxE interactions for environmental stresses that endangered species are expected to encounter with greater frequency and magnitude in the future, e.g. by selecting stresses based on well-established predictive models of regional climate change. CGV that is revealed under such conditions should provide a more adaptively significant measurement of biodiversity than existing measures of intraspecific genetic variation and thus a more valuable metric for determining what individuals and lineages are important to species conservation efforts. With limited conservation budgets, preserving this much smaller CGV footprint should provide a more realistic and realizable conservation goal. Similar principles can be extended to transform existing measures of species richness into more contextual measures of species response diversity [43] that are anticipated to be most relevant to future ecological resilience.

**Cancer:** The same principles for preserving populations can also be used in their eradication. In highly evolvable infections and diseases such as aggressive cancers, the accumulation of CGV is likely to play an important role in the rapid evolution of therapy resistance that is seen in the latest generation of targeted therapies [32]. Importantly, oncologists have yet to consider strategies for reducing tumor evolvability by eliminating CGV. The arguments reviewed here suggest that environment stability facilitates CGV accumulation in tumors and plays an important role in the evolution of therapy resistance. Therapeutic strategies that rapidly change the administered therapy over time should help to eliminate CGV and thereby provide a novel strategy for reducing the evolutionary potential of cancers that contain high levels of genetic diversity [44].

**Artificial Evolution in Molecular Systems:** Degeneracy could also be exploited within artificial molecular systems such as modified polymerase chain reaction (PCR). In previous work we have shown that PCR primers bind to the target templates degenerately and compete for targets [45]. Primers could be designed and tested that are different in DNA sequence but bind to similar targets, modeling CVG. Subsequent experiments could test amplification of new targets by these primers, thereby modeling a previously static population placed in a new environment. The arguments outlined in this review suggest that primers with broader degeneracy in annealing to diverse templates would amplify targets faster compared to more specific templates, unless primers with narrow specificities happen, by chance, to be highly specific for new templates. In the latter case, "adaptation" of primers to new templates would happen due to diversity, not degeneracy of the primer population.

**Adaptation in Socio-Technical Systems:** The persistence and spreading of technological and cultural artifacts through time with variation is analogous to variation with heritability in that both are examples of what Darwin called "descent with modification" [46]. Biased (non-random) selection can be imposed on both types of processes and consequently both can undergo a form of evolution. While nature appears to have largely resolved conflicts between short-term (stabilizing) and long-term (directional) selection, this cannot be said of the short and long-term objectives that arise in the planning and operations of adapting socio-technical systems or in systems engineering. While CGV is a purely biological phenomenon, the broader concept of degeneracy captures system properties that can be clearly articulated and defined for any system comprised of functionally versatile elements. Currently, there are several research programs that are exploring how the degeneracy concept can be translated into design principles for the realization of more flexible and resilient systems in several disciplines [4, 5, 47, 48]. For instance, in Defense capability studies, we have shown using simulations that fleets of land field vehicles with high degeneracy in task capabilities can improve operational robustness within anticipated mission scenarios yet at the strategic level provides exceptional design and organizational adaptability for responding to unanticipated challenges [5]. More than just an academic exercise, this role of degeneracy in adaptation has attracted the interest and financial support of Australian Defense. Consequently, these principles will be presented this year at NATO's strategic planning conference as a proposal for mitigating strategic uncertainty [49]. Because the degeneracy concept is itself very versatile, we are looking at how this concept can also be translated in the design of more flexible manufacturing and assembly systems [47], and for better performance in population-based dynamic optimization [4]. Still others are using these concepts to understand some of the weaknesses of contemporary peer review processes [6] and the requisite conditions for embodied [50] and simulated artificial life [51-53].

As the similarities between degeneracy and CGV become better appreciated and their fundamental role in evolution becomes more widely accepted, we anticipate that these principles will transform how we think about evolutionary processes and the origins of innovation.

# List of abbreviations

CGV, cryptic genetic variation; GxE, genotype-environment interactions.

# Competing interests

The authors declare that they have no competing interests.

Draft: Submitted to Integrative Biology (2011)

# Acknowledgements

We are grateful for insightful comments and suggestions from Andreas Wagner, Axel Bender, Angus Harding, and the reviewers from the Biology Direct editorial board.

# Authors' contributions

Both authors contributed to the concept and manuscript preparation.

# Funding

Dr. Atamas' research is funded by the NIH R21 HL106196 and VA Merit Review Award
Dr. Whitacre's research is funded by an Australian DSTO grant